\documentclass[a4paper]{jpconf}
\usepackage{graphicx,amsmath,cite,lscape}
\begin{document}
\title{Gravitational-wave standard siren without redshift identification}

\author{Atsushi Nishizawa$^1$, Kent Yagi$^2$, 
Atsushi Taruya$^{3,4}$, and Takahiro Tanaka$^1$}

\address{$^1$ Yukawa Institute for Theoretical Physics, Kyoto University, 
Kyoto 606-8502, Japan}
\address{$^2$ Department of Physics, Kyoto University, 
Kyoto 606-8502, Japan}
\address{$^3$ Research Center for the Early Universe, Graduate School of 
Science, The University of Tokyo, Tokyo 113-0033, Japan}
\address{$^4$ Institute for the Physics and Mathematics of the Universe, 
The University of Tokyo, Kashiwa, Chiba 277-8568, Japan}

\ead{anishi@yukawa.kyoto-u.ac.jp}

\begin{abstract}
Proposed space-based gravitational-wave (GW) detectors such as DECIGO and BBO 
will detect $\sim10^6$ neutron-star (NS) binaries and determine the luminosity distances to the binaries with high precision. Combining the luminosity distances with cosmologically-induced phase corrections on the GWs, cosmological expansion out to high redshift can be measured without the redshift determinations of host galaxies by electromagnetic observation and can be a unique probe for dark energy. This article is based on the results obtained in \cite{bib99} where we investigated constraining power of the GW standard siren without redshift information on the equation of state of dark energy with future space-based GW detectors. We also compare the results with those obtained with other instruments and methods.
\end{abstract}

\section{Introduction}
\hspace*{\parindent}Future space-based gravitational-wave (GW) detectors such as DECI-hertz Interferometer Gravitational-wave Observatory (DECIGO) \cite{bib1,bib2} and Big-Bang Observer (BBO) \cite{bib3} (see also \cite{bib4} for updated information) are the most sensitive to GWs in $0.1 - 1\,{ \rm{Hz}}$ band and will aim at detecting the primordial GW background, the mergers of intermediate-mass black holes (BH), and a large number of neutron-star (NS) binaries in an inspiraling phase. 

It is known that the continuous GW signal from a compact-binary object provides a unique way to measure the luminosity distance to the source with high precision. Such binary sources are often referred to as the standard siren. With the redshift information determined by an electromagnetic (EM) follow-up observation, the standard siren can be an accurate tracer of the cosmic expansion \cite{bib6}. The potential power of this method as a dark-energy probe has been investigated with ground- and space-based detector configurations for GWs \cite{bib4,bib21,bib22,bib23,bib24,bib25,bib26,bib27,bib28,bib88,bib37,bib99, bib98}. The most of the preceding works assume that the redshifts of all GW sources are known by EM observations. However, this assumption is rather strong and is too optimistic to be justified because the spectroscopic follow-up observation of galaxies is, in general, time-consuming, particularly at high redshifts \cite{bib84}. In addition, not all galaxies can be observed due to intrinsic faintness, the absence of spectral features, limited sky coverage, and a limited redshift range (redshift desert). As a result, in a practical follow-up observation, the fraction of the binary sources whose redshifts are spectroscopically obtained is significantly reduced. From a rough estimate based on the number density of galaxies potentially observable and the number of galaxies to be observed in the future galaxy redshift surveys such as JDEM/WFIRST \cite{bib86} and Euclid \cite{bib87}, it turns out that the fraction of redshift identification is $\sim 10^{-4}$ with uncertainty of about one order of magnitude \cite{bib99}. This means that the GW sources with redshift information are quite rare unless we perform a large-scale follow-up campaign dedicated for GW events.

Another approach is to measure the cosmic-expansion rate from GW observations alone without electromagnetically-estimated redshifts. As suggested in \cite{bib1}, the cosmic expansion affects not only the amplitude (luminosity distance) of GWs but also the phase, and it is possible to directly obtain information about the cosmic acceleration by accurately measuring the GW phase shift of a binary source at a certain redshift. Although the redshifts of the binaries are assumed to be determined by the EM follow-up observations in the previous works \cite{bib1,bib20}, if we combine the independent information from the luminosity distance and the cosmological phase shift, we can measure the expansion history of the universe without any reference to the EM counterpart or host-galaxy identification.  Furthermore, this method based on a purely GW observation enables us to compare the observational data with those obtained in other EM observations. Since proposed space-based GW detectors such as DECIGO and BBO would detect $\sim10^6$ NS binaries, they provide a novel opportunity to measure the property of dark energy without using the cosmic ladders. 

Our primary interest here is in the potential of the standard siren without any help of redshift information of the binary sources and how the sensitivity is improved by partially adding the redshift information of a fraction of NS binaries. We show the constraining power of the GW standard siren with/without redshift information on the equation of state (EOS) of dark energy. As GW detectors, we use future space-based detector, DECIGO and BBO, allowing their noise curves to be scaled appropriately and including confusion noises produced by astrophysical sources. Throughout the paper, we adopt units $c=G=1$.

\section{Detector noise and astrophysical foregrounds}
\label{sec2}
\hspace*{\parindent}In this section, we summarize the detector-noise curves of DECIGO and BBO and confusion noises from astrophysical GW foregrounds.

DECIGO and BBO are the most sensitive to GWs in $0.1 - 1\,{\rm{Hz}}$ band \cite{bib1,bib3}. Using the currently-proposed design parameters \cite{bib2}, we obtained fitting formula for the sky-averaged noise curve of DECIGO single interferometer: 
\begin{equation}
\hat{S}_{\rm{h,\,D}}^{\rm{inst}} (f) = 3.30\times 10^{-50} \biggl(\frac{f}{1\rm{Hz}} \biggr)^{-4} +3.09\times 10^{-47} \left[1+\left(\frac{f}{f_{\rm{c}}}\right)^2 \right] \;\; 
\rm{Hz}^{-1} \;, 
\label{eq4}
\end{equation}
where $f_{\rm{c}}=7.69\,{\rm{Hz}}$. BBO has a slightly different noise shape from DECIGO, because of the different interferometer type and optical parameters. In the presence of WD confusion noise below $\sim 0.1 \,{\rm{Hz}}$, the BBO sensitivity to a NS binary is nearly three times better in amplitude than that of DECIGO.

DECIGO and BBO are sensitive to a large number of astrophysical GW sources, in particular NS, BH, and WD binaries. It is expected that the low-frequency side of the noise curve would be dominated by the GWs from extra-galactic population of WD binaries, which remains after subtracting individually identified signals and are stochastic in nature. According to the estimation by \cite{bib7}, the fitting formula for the power spectrum of WD confusion noise \cite{bib12} (residual contribution after the subtraction) is given by
\begin{equation}
S_h^{\rm{WD}}(f) = 4.2 \times 10^{-47} \left( \frac{f}{1\,{\rm{Hz}}} \right)^{-7/3}\, \exp \left[-2 \left( \frac{f}{5\times 10^{-2}\,{\rm{Hz}}} \right)^2 \right] \;\; {\rm{Hz}}^{-1}\,\;. \nonumber
\end{equation}
There is also contribution from galactic WD binaries \cite{bib12}. However, we verified that the galactic contribution is negligible above the frequency $5\times 10^{-3}\,{\rm{Hz}}$.

Another astrophysical source that we have to take into account is the NS-binary foreground. According to \cite{bib14}, the energy density of GWs from NS binaries per logarithmic frequency bin normalized by the critical energy density of the universe at present is written as
\begin{equation}
\Omega_{\rm{gw}}^{\rm{NS}} (f) = \frac{8 \pi^{5/3}}{9 H_0^2} M_c^{5/3} f^{2/3} n_0 \;, \quad \quad n_0 = \int_0^{\infty} \frac{\dot{n}(z)}{(1+z)^{4/3} H(z)} dz \;. 
\label{eq3}   
\end{equation}
$H_0$ is the Hubble constant and $M_c$ is the chirp mass defined as $M_c \equiv \eta^{3/5} M_t$, together with the total mass $M_t=m_1+m_2$ and the symmetric 
mass ratio $\eta=m_1 m_2/M_t^2$. $\dot{n}(z)$ is the NS merger rate per unit comoving volume per unit proper time at a redshift $z$. We adopt the following fitting form of the NS-NS merger rate 
given in \cite{bib15}: $\dot{n}(z) = \dot{n}_0\, s(z)$, $s(z) =1+2 z$ for $z \leq 1$, $\frac{3}{4} (5-z)$ for $1 < z \leq 5$, $0$ for $5 < z$. This function $s(z)$ is estimated based on the star formation history inferred from the UV luminosity \cite{bib16}. 
The quantity $\dot{n}_0$ represents the merger rate at present. By assuming the flat $\Lambda$CDM universe with $\Omega_{\rm{m}}=0.3$ ($\Omega_{\Lambda}=0.7$) and the Hubble constant $H_0=h_{72} \times 72\,{\rm{km}}\,{\rm{s}}^{-1}\,{\rm{Mpc}}^{-1}$, we obtain the NS confusion-noise power spectrum 
\begin{equation}
S_h^{\rm{NS}} (f) = 1.55 \times 10^{-47} h_{72}^{-1} \left( \frac{M_c}{1.22 M_{\odot}} \right)^{5/3} \left( \frac{f}{1\,{\rm{Hz}}} \right)^{-7/3} \left( \frac{\dot{n}_0}{10^{-6}\,{\rm{Mpc}}^{-3}\,{\rm{yr}}^{-1}} \right) \;, \nonumber 
\end{equation}
where we used the relation $S_{\rm{h}}^{\rm{NS}}(f) = (3H_0^2/4\pi^2 f^3 ) \Omega_{\rm{gw}}^{\rm{NS}}(f)$ \cite{bib19}.

\begin{figure}[t]
\begin{center}
\includegraphics[width=7cm]{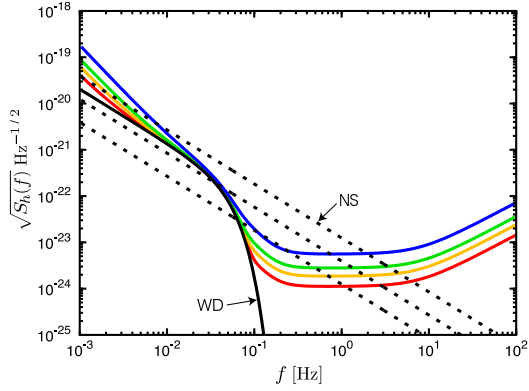}
\caption{Noise curves of default and scaled DECIGO: $r_n=1$ (blue), $1/2$ (green), $1/3$ (orange), and $1/5$ (red) from the top to the bottom. These noise curves include the confusion noise from a number of WD binaries. The WD-binary foreground $\sqrt{S_h^{\rm{WD}}}$ is shown with solid, black curve on the left side. The three diagonal dashed lines represent the NS-binary foreground $\sqrt{S_h^{\rm{NS}}}$ before subtraction (or ${\cal{R}}_{\rm{NS}}=1$) with $\dot{n}_0=10^{-5},\,10^{-6},\,10^{-7}\,{\rm{Mpc}}^{-3}\,{\rm{yr}}^{-1}$ from the top to the bottom.}
\label{fig2}
\end{center}
\end{figure} 

Including astrophysical contributions, a total-noise power spectrum is given by 
\begin{equation}
S_h(f)=S_h^{\rm{inst}}(f) + S_h^{\rm{WD}}(f)+ S_h^{\rm{NS}}(f) {\cal{R}}_{\rm{NS}} \;. 
\label{eq6}
\end{equation}
Since our purpose here is to assess cosmology achieved by space-based detectors such as DECIGO and BBO and to make clear the requirement for the experimental design, we consider only DECIGO, but allowing the noise curve in Eq.~(\ref{eq4}) to vary by an overall factor $r_n=1$, $1/2$, $1/3$, $1/5$ in amplitude. Namely, $S_h^{\rm{inst}} (f) = r_n^2 \times \hat{S}_{\rm{h,\,D}}^{\rm{inst}} (f)$. Note that BBO noise power spectrum approximately corresponds to that of DECIGO with $r_n=1/3$. The factor ${\cal{R}}_{\rm{NS}}$ denotes a suppression factor due to the subtraction of individually identified NS binaries, whose value strongly depends on the detector noise curve. According to \cite{bib99} in which the authors have fitted the numerical result obtained by Yagi and Seto \cite{bib17}, the suppression factor is ${\cal{R}}_{\rm{NS}}=0$ except for $r_n=1$ case. When $r_n=1$, some amount of  the foreground residuals is left and gives ${\cal{R}}_{\rm{NS}}\approx 4.6 \times 10^{-3}$. The contribution of each term in Eq.~(\ref{eq6}) is shown in Fig.~\ref{fig2}.

\section{Standard siren as a probe for dark energy}
\label{sec5}

\subsection{Standard siren}
\label{sec3}
\hspace*{\parindent}The continuous GW signal from a compact-binary object is often 
referred to as standard sirens \cite{bib6}. For a single binary system, the Fourier transform of the GW waveform is 
expressed as a function of frequency $f$ \cite{bib18,bib19},
\begin{equation}
\tilde{h} (f) = \frac{A}{d_L(z)} M_z^{5/6} f^{-7/6} e^{i \Psi(f)} \;, 
\label{eq9}
\end{equation}
where $d_L$ is the luminosity distance, and  
the quantity $M_z=(1+z) M_c$ is the 
redshifted chirp mass. $M_c$ is the proper chirp mass defined in the source rest frame. The constant $A$ is given by $A= (\sqrt{6}\, \pi^{2/3})^{-1}$, 
which includes the factor $\sqrt{4/5}$ for a geometrical 
average over the inclination angle of a binary \cite{bib80}. The function $\Psi(f)$ represents the frequency-dependent phase arising from 
the orbital evolution, and at the order of the {\it{restricted}} $1.5$ 
post-Newtonian (PN) approximation, it is given by \cite{bib18,bib19} 
\begin{align}
\Psi (f) &= 2\pi f \,t_c -\phi_c -\frac{\pi}{4} +\frac{3}{128} 
(\pi M_z f)^{-5/3} \nonumber \\
& \times \left[ 1+ \frac{20}{9} \left(\frac{743}{336} + 
\frac{11}{4} \eta \right)
\eta^{-2/5} (\pi M_z f)^{2/3} - 16 \pi \eta^{-3/5} (\pi M_z f) - \frac{25}{768} X(z) M_z (\pi f M_z)^{-8/3} \right] \;, 
\label{eq8}
\end{align}
where $t_c$ and $\phi_c$ are the time and phase at coalescence, 
respectively. The last term in the bracket is phase correction due to cosmic expansion ($-4$ PN order) \cite{bib1,bib20}, where $X(z)$ is defined as $X(z) \equiv \frac{1}{2} [ H_0 -H(z)/(1+z)]$ or equivalently expressed as $[ \dot{a}(0)-\dot{a}(z) ]/2$. Thus, $X(z)>0$ ($X(z)<0$) corresponds to the accelerating (decelerating) universe.  

There are various methods of the standard siren. The simplest minimal one is to measure the amplitude of GWs, or the luminosity distance as a function of redshifts. However, seen from Eq.~(\ref{eq9}), GW observation itself cannot provide a source redshift, since the mass parameter we can determine is $M_z=(1+z)M_c$ and the source redshift degenerates with the proper source mass $M_c$. So the redshift has to be determined from EM observation of host galaxies. This method as a tool to measure the cosmic expansion has been investigated by many authors \cite{bib4,bib21,bib22,bib23,bib24,bib25,bib26,bib27,bib28}, assuming that the source redshifts are known by spectroscopic follow-up observations. Even in the situation that the redshift of a source is not uniquely determined due to multiple candidates of a host galaxy in a detector error cube, one can infer the most likely cosmological model by statistically analyzing large samples of observed events \cite{bib6,bib94}. However, to do that, one needs a complete catalog of galaxy redshifts. This assumption is rather strong and is too optimistic to be justified, because of the smallness of the fraction of redshift identification ($\sim10^{-4}$) \cite{bib99}. 

To break the redshift degeneracy and measure the cosmic expansion by GW observation alone, recently some elaborate methods have been proposed. The redshift degeneracy could be broken if we know the EOS of a NS {\it{a priori}} \cite{bib89} or if the observed chirp mass distribution of NS is sufficiently narrow \cite{bib92}. We will discuss the sensitivities of several methods later in Sec.~\ref{sec4}.

In this article, we investigate another method breaking the redshift degeneracy by the phase modulation of GWs due to the cosmic expansion, which is the last term including $X(z)$ in Eq.~(\ref{eq8}) \cite{bib1,bib20}. Although in the preceding works the redshifts of the binaries are assumed to be determined by the EM follow-ups \cite{bib1,bib20}, the redshift information in principle is not mandatory if independent information from both the luminosity distance and the phase shift is available. In fact, as indicated by our results shown later, we can measure a cosmic-expansion rate only with GW observations without electromagnetically-estimated redshifts. This is a great advantage of this method, because a pure GW observation enables us to compare its observational data with those obtained in other EM observations. 

In the subsequent sections, we investigate the power of utilizing the phase correction without assuming any electromagnetically-estimated redshift, and give the figure of merits. Then we compare this method with the previous one, in which source redshifts are given, and combine both methods to maximize the accuracy of the estimation.

\subsection{Numerical estimation of measurement accuracy}
\label{sec5b}
\hspace*{\parindent}We consider a spatially flat universe with dark energy whose EOS is parameterized as $w(z) = w_0 +w_a z/(1+z)$. The parameters to be determined from GW observations are $w_0$, $w_a$, $\Omega_m$, and $H_0$. For the analysis of error estimation, we adopt a fiducial set of cosmological parameters: $w_0=-1$, $w_a=0$, $\Omega_m=0.3$, $H_0=72\,{\rm{km}}\,{\rm{s}}^{-1}\,{\rm{Mpc}}^{-1}$.   

In what follows, we calculate the estimation errors of binary parameters $\theta _a$: $M_z$, $\eta$, $t_c$, $\phi_c$, $d_L$, and $X$ in the waveform of Eq.~(\ref{eq9}) with the Fisher information matrix. It is well known that the Fisher matrix analysis underestimates the parameter errors if SNR is not sufficiently large. Even default DECIGO (8 identical interferometers) is able to observe a NS binary at $z=1$, $3$, and $5$ with SNR about 24, 31, and 67, respectively. Therefore, it is justified to use the Fisher matrix for the parameter estimation. The Fisher matrix for a single binary 
is given by \cite{bib18,bib29}
\begin{align}
\Gamma_{ab}^{({\rm{single}})} &= 4 \sum_{i=1}^{8}\, {\rm{Re}} \int_{f_{\rm{min}}}^{f_{\rm{max}}}
 \frac{\partial_{a} \tilde{h}^{\ast}(f)\, \partial_{b}
 \tilde{h}(f)}{S_{\rm{h}}(f)} df \;,
 \label{eq13}
\end{align}
where $\partial_a$ denotes a derivative with respect to a parameter $\theta_a$. Since the 8 interferometers of DECIGO and BBO are assumed to be identical, the summation is reduced to just multiplying a factor 8. The noise power spectrum and GW signal are given by Eqs.~(\ref{eq6}) and (\ref{eq9}), respectively. The lower cutoff frequency $f_{\rm min}$ is given by, $f_{\rm{min}} = 0.233 \left( 1M_{\odot}/M_z \right)^{5/8} 
\left( 1\,{\rm{yr}}/T_{\rm{obs}} \right)^{3/8} \; {\rm{Hz}}$,
and the upper cutoff of the frequency is set to $f_{\rm{max}}=100\,{\rm{Hz}}$ which naturally arises from the upper cutoff of the noise curve. Given the numerically evaluated Fisher matrix, 
the marginalized 1-$\sigma$ error of a parameter, $\Delta\theta_a$,
is estimated from the inverse Fisher matrix, $\Delta \theta_a = \sqrt{\{\mathbf{\Gamma}^{-1}\}_{aa}}$.

Once the estimation error of $X$ is obtained, it is straightforward to calculate the estimation errors of the cosmological parameters. Since the error of $d_L$ is much smaller than that of $X$ (see Appendix of \cite{bib99}), we can replace observed $d_L$ with the corresponding redshift in the fiducial cosmological model when we derive the measurement accuracies of cosmological parameters from $X$. Thus, for the simplicity of the analysis, we use $X(z)$ instead of $X(d_L)$. Furthermore, we assume that the Hubble constant $H_0$ is known {\it{a priori}}, and fix it to $H_0=72\,{\rm{km}}\,{\rm{s}}^{-1}\,{\rm{Mpc}}^{-1}$, because the Hubble constant has been determined at a-few-percent level from the observation of nearby Cepheids and supernovae \cite{bib32}. Thus, the free parameters of the Fisher matrix are $w_0$, $w_a$, and $\Omega_m$. The Fisher matrix is given by
\begin{equation}
\Gamma_{ab} = \int_{0}^{\infty} \frac{\partial_a X(z) \partial_b X(z)}{\sigma_X^2(z)} \frac{d N(z)}{dz} dz\;, \quad \frac{dN(z)}{dz}=T_{\rm obs}\,\frac{4\pi r^2(z)}{H(z)}\frac{\dot{n}(z)}{1+z} \;, 
\label{eq12}
\end{equation}
where $dN(z)/dz$ is the number of NS binaries in the redshift interval $[z,z+dz]$ observed during $T_{\rm obs}$ \cite{bib15}. $r(z)$ is the comoving radial distance defined as $r(z)=d_L(z)/(1+z)$ and $\dot{n}$ is given in Sec.~\ref{sec2}. Since the normalization of $\dot{n}$ is still uncertain, we adopt the most recent estimate \cite{bib30}, $\dot{n}_0=10^{-6}\,{\rm{Mpc}}^{-3}\, {\rm{yr}}^{-1}$. Since the observation time $T_{\rm{obs}}$ is a crucial parameter, we suppose 3-yr, 5-yr, and 10-yr observation.

\subsection{Results}
\begin{figure*}[t]
\begin{center}
\includegraphics[width=12cm]{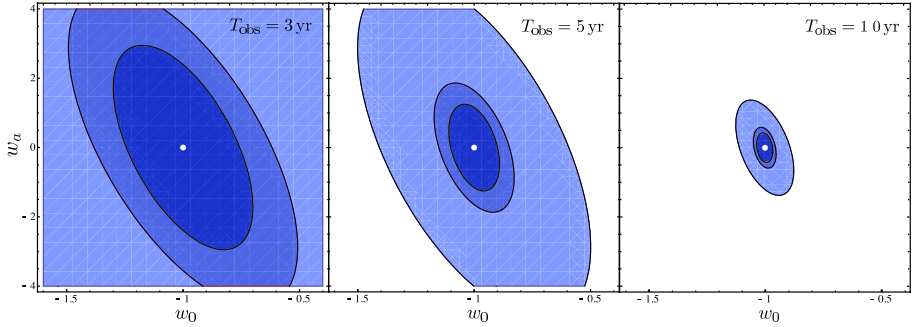}
\caption{$w_0$ - $w_a$ error ellipses marginalized over $\Omega_m$. The merger rate is fixed to $\dot{n}_0=10^{-6}\,{\rm{Mpc}}^{-3}\,{\rm{yr}}^{-1}$. From the left to the right panels, the observation time is $3\,{\rm{yr}}$, $5\,{\rm{yr}}$, and $10\,{\rm{yr}}$. In each panel, the larger to smaller ellipses denote those with $r_n=1$, $1/3$, and $1/5$, respectively. The dot at the center is our fiducial value: $w_0=-1$ and $w_a=0$.}
\label{fig6}
\end{center}
\end{figure*}

\hspace*{\parindent}The measurement accuracies of $w_0$ and $w_a$ marginalized over the other remaining parameters are shown in Fig.~\ref{fig6} for the fixed $\dot{n}_0$  varying the values of $r_n$ and $T_{\rm{obs}}$. The size of error ellipses change significantly depending on the parameters $T_{\rm{obs}}$ and $r_n$. For example, if we choose $T_{\rm{obs}}=5\,{\rm{yr}}$, the errors $\Delta w_0$ and $\Delta w_a$ are 0.330 and 3.254 for $r_n=1$, and 0.113 and 1.231 for $r_n=1/3$. For 10-yr observation, $\Delta w_0$ and $\Delta w_a$ reach at the levels of several $\times 10^{-2}$ and a few $\times 10^{-1}$, respectively. The more comprehensive parameter surveys are found in \cite{bib99}.

To show the results more clearly, let us define the figure of merit (FoM) for dark energy, ${\rm{FoM}} \equiv \sqrt{\det \gamma_{ab}}$, where $\gamma_{ab}$ is the inverse matrix of $(\Gamma^{-1})_{ab}$ with $a,b=w_0,\,w_a$. This FoM is inversely proportional to the area of an error ellipse on the $w_0$ - $w_a$ plane 
and ranges from $\sim 10^{-1}$ to $\sim 10^3$. Since DECIGO and BBO will be launched in the late 2020s, the FoM should be compared with other future projects of an EM observation probing dark energy at that time: type-Ia supernovae, baryon acoustic oscillation, or weak lensing surveys. So typical criteria in FoM would be from 10 to 100, which correspond to future projects of stage III and IV in the dark energy task force \cite{bib71}, respectively. To achieve these criteria, observation time longer than $5\,{\rm{yr}}$ is preferable. Since the FoM rapidly improves being roughly proportional to $T_{\rm{obs}}^{31/8}$ \cite{bib99}, the observation time is a crucial factor. Given 10-yr observation and the typical rate of binary mergers, the FoM is $\sim 100$ and is comparable to a stage-IV project in the dark energy task force. It should be emphasized that this method requires no redshift information of the GW sources and is completely independent of any EM observation.

\subsection{Adding redshift information of binaries}
\hspace*{\parindent}In the previous subsections, we did not assume any redshift information to measure the dark-energy parameters. Although it would be too ideal to assume that all binary redshifts are well determined by EM follow-up observations, some fraction of them will be determined by the follow-up observations. Then, we can optimize the constraints on the dark-energy parameters by using thus-determined redshifts. In this subsection, we combine $z$ - $d_L$ and $d_L$ - $X$ information and evaluate what fraction is needed to achieve ${\rm{FoM}}=100$, which is a typical value expected in future dark energy surveys \cite{bib71}. 

The relevant Fisher matrix is defined with the redshift-identification fraction $\alpha$ by
\begin{equation}
\Gamma_{ab} = \alpha \int_0^{\infty} \frac{\partial_a d_L (z) \partial_b d_L(z)}{\sigma_{d_L}^2(z)} \frac{d N(z)}{dz} dz + \int_0^{\infty} \frac{\partial_a X(z) \partial_b X(z)}{\sigma_X^2(z)} \frac{d N(z)}{dz} dz \;, 
\label{eq14}
\end{equation}
where $\sigma_{d_L}$ is an error in the luminosity distance. Strictly speaking, the variables $d_L$ and $X$ cannot be treated independently as in Eq.~(\ref{eq14}). However, the correlation coefficient between $d_L$ and $X$ is of the order of $10^{-6}$. Hence, we can safely use the expression in Eq.~(\ref{eq14}). The fraction factor $\alpha$ would be a function of a redshift in a real galaxy survey, but we take it as a constant for simplicity of the analysis.

We estimate the error size of the luminosity distance via the same procedure as in Sec.~\ref{sec5b}, but including possible systematic errors: weak-lensing magnification due to the matter inhomogeneities along the line of sight \cite{bib34} and the Doppler effect due to the random peculiar velocities of binary sources \cite{bib36}. 
These systematic errors to the luminosity distance are summarized as 
\begin{align}
\sigma_{d_L}^2 (z) &\equiv \left[ \frac{\Delta d_L(z)}{d_L(z)} \right]^2 = 
\sigma_{\rm{inst}}^2(z) + \sigma_{\rm{lens}}^2(z) + \sigma_{\rm{pv}}^2(z) \;,
\label{eq22} \\
\sigma_{\rm{lens}}(z) &=0.066 \left[ \frac{1-(1+z)^{-0.25}}{0.25} \right]^{1.8} 
\;, \quad \quad \sigma_{\rm{pv}}(z) = \left| 1-\frac{(1+z)^2}{H(z)d_L(z)} \right| \sigma_{\rm{v,gal}} \;, \nonumber 
\end{align}
where $\sigma_{\rm{inst}}$, $\sigma_{\rm{lens}}$, and $\sigma_{\rm{pv}}$ are induced by the instrumental noise, the lensing magnification, and the peculiar velocity of binaries, respectively. $\sigma_{\rm{v,gal}}$ is the one-dimensional velocity dispersion of the galaxy and is set to be $\sigma_{\rm{v,gal}}=\,300$km\,s$^{-1}$. For the plot of the redshift dependence of each error term and the details, see \cite{bib37}. The important thing is that the lensing error dominates at almost all redshift range.

\begin{figure}[t]
\begin{center}
\includegraphics[width=7cm]{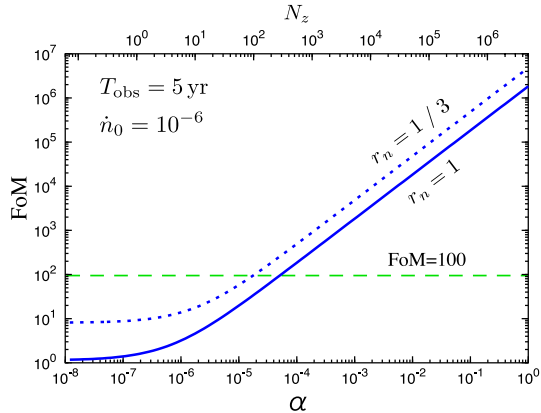}
\caption{FoM as a function of $\alpha$ defined in Eq.~(\ref{eq14}) or $N_z$. The observation time and the merger rate are fixed to $T_{\rm{obs}}=5\,{\rm{yr}}$ and $\dot{n}_0=10^{-6}\,{\rm{Mpc}}^{-3}\,{\rm{yr}}^{-1}$. The horizontal dashed line represents ${\rm{FoM}}=100$. The solid and dotted curves are the case with $r_n=1$ and $1/3$, respectively.}
\label{fig11}
\end{center}
\end{figure} 

Then using Eq.~(\ref{eq14}), we estimate the errors of $w_0$ and $w_a$ and corresponding FoM as a function of $\alpha$. We also calculate the corresponding number of NS binaries $N_z$ whose redshifts are identified. The result is shown in Fig.~\ref{fig11} for typical values of parameters, $T_{\rm{obs}}=5\,{\rm{yr}}$ and $\dot{n}_0=10^{-6}\,{\rm{Mpc}}^{-3}\,{\rm{yr}}^{-1}$. At low $\alpha$, of course, the $z$ - $d_L$ information plays no significant role in improving the FoM. But at $\alpha \approx$ a few $\times10^{-6}$, $z$ - $d_L$ information begins to contribute to FoM, and enables the FoM to reach 100 at around $\alpha \approx$ a few $\times 10^{-5}$. As anticipated, $z$ - $d_L$ measurement is much more powerful than the $d_L$ - $X$ measurement because ${\rm{FoM}}|_{\alpha=0}$ is much smaller than ${\rm{FoM}}|_{\alpha=1}$. Therefore, identifying the source redshifts strongly assists the $d_L$ - $X$ measurement.

One may concern how much fraction $\alpha$ is feasible in the future galaxy redshift surveys. From our rough estimate \cite{bib99} based on the number density of galaxies potentially observable and the number of galaxies to be observed in the future galaxy redshift surveys such as JDEM/WFIRST \cite{bib86} and Euclid \cite{bib87}, it turns out that the fraction of redshift identification is $\sim10^{-4}$ with uncertainty of about one order of magnitude. This means that the GW sources with redshifts are quite rare unless we perform a large-scale follow-up campaign dedicated for GW events. So it turns out that $d_L$ - $X$ measurement may be more sensitive than $z$ - $d_L$ measurement and that the standard sirens without redshifts guarantee minimally achievable FoM.  

\section{Comparison with previous results with other detectors and methods}
\label{sec4}

\begin{table}[t]
\caption{\label{label}List of cosmological constraints obtained with other detectors and methods. Here we list only the detectors that can probe at cosmological distance and works that take into account weak lensing error. SMBH and CMB refer to supermassive black hole and cosmic microwave background, respectively.}
\begin{center}
\begin{tabular}{llllll}
\br
reference & detector &GW       & number   & source redshift &  method to  \\
                &               & source & of events  &                        & determine $z$ \\
\mr
Van Den Broeck et al. \cite{bib27} & LISA & SMBH & 1 & host galaxy & identifying \\
& & & & & a single source \\
Petiteau et al. \cite{bib88} & LISA & SMBH & $\sim40$ & host galaxy & statistical \\
& & & & & \\
Sathyaprakash et al. \cite{bib26} & ET & NS & $10^3$ & EM counterpart & identifying  \\
& & & & (short GRB) & a single source \\
Zhao et al. \cite{bib28} & ET & NS & $10^3$ & EM counterpart & identifying \\
& & & & (short GRB) & a single source \\
Messenger \& Read \cite{bib89} & ET & NS & N/A & unnecessary & tidal effect \\
& & & & & on GW phase  \\
Cutler \& Holz \cite{bib4} & DECIGO & NS & $\sim10^5$ & host galaxy & identifying \\
& or BBO & & & & a single source \\
Nishizawa et al. \cite{bib99} & DECIGO & NS & $\sim10^6$ & unnecessary & cosmological \\
(this work) & or BBO & & & & phase drift \\
\br
\end{tabular}

\begin{tabular}{lllll}
\br
reference & free              & fixed            & prior &  sensitivity \\
                & parameters & parameters &           &                \\
\mr
Van Den Broeck et al. \cite{bib27} & 
$w_0$ &  $w_a$, $\Omega_m$, & none & $\Delta w_0 \sim 0.2$ \\
& & $\Omega_k$, $H_0$ & & \\
Petiteau et al. \cite{bib88} & $w_0$ & $w_a$, $\Omega_m$, & CMB & $\Delta w_0 \approx 0.036$ \\
& & $\Omega_k$, $H_0$ & (WMAP) & \\
Sathyaprakash et al. \cite{bib26} &
$w_0$, $\Omega_m$, $\Omega_k$,  & $w_a$ & none & $\Delta w_0 \approx 0.23$ \\
 & $H_0$ & & & \\
Zhao et al. \cite{bib28} &
$w_0$, $w_a$, $\Omega_m$, & none & CMB & $\Delta w_0 \approx 0.099$ \\
 & $\Omega_k$, $H_0$ & & (Planck) & $\Delta w_a \approx 0.302$ \\
Messenger \& Read \cite{bib89} & N/A & N/A & NS EOS & $\Delta z/z \sim 0.1$ \\
& & & & (for an event) \\
Cutler \& Holz \cite{bib4} &
$w_0$, $w_a$, $\Omega_m$, & $\Omega_k$ & CMB & $\Delta w_0 \sim 0.01$ \\
 & $H_0$ & & (Planck) & $\Delta w_a \approx 0.1$ \\
Nishizawa et al. \cite{bib99} & $w_0$ $w_a$, $\Omega_m$ & $\Omega_k$, $H_0$ & none & $\Delta w_0 \approx 0.113$ \\
(this work) & & & & $\Delta w_a \approx 1.23$ \\
\br
\end{tabular}
\end{center}
\label{tab1}
\end{table}

\hspace*{\parindent}Preceding works for other detectors are listed in Table \ref{tab1}, in which only the detectors that can probe at cosmological distance are shown. As for DEICGO/BBO, our results should be compared with the work by Cutler and Holz \cite{bib4}. They assume that one can identify a single host galaxy due to the good angular resolution of DECIGO/BBO and that the redshift of the host galaxy is always obtained by an EM follow-up observation. However, as estimated in \cite{bib99}, the assumption is too optimistic to be justified, because the redshift identification of host galaxies is time-consuming and inefficient and whose fraction would be at worst $\sim10^{-4}$ even for the future galaxy surveys such as JDEM/WFIRST and Euclid. Thus, our results are conservative because we do not require any redshift identification and any prior from other observations. We can conclude that the standard siren without source redshifts guarantee minimally achievable FoM, which can be improved by increasing the fraction of the galaxy redshift identification and combining other EM observations.

One may also compare the results in this paper with those obtained from Einstein Telescope (ET). The total number of NS binaries observed by ET is comparable to DECIGO. However, ET has to rely on short gamma-ray bursts (GRB) to determine the source redshift so that the number of sources with redshifts is limited by short GRB directed toward us and is reduced to $\sim10^3$ or smaller. Consequently, the sensitivity of DECIGO/BBO without redshift information is twice better than that of ET by Sathyaprakash et al. \cite{bib26}. If the Planck prior of CMB is added as done by Zhao et al. \cite{bib28}, the size of the error is significantly reduced. On the other hand, the method proposed by Messenger and Read \cite{bib89} does not assume source redshifts. However, to break the parameter degeneracy between the chirp mass and redshift of a NS binary, one needs to assume the EOS of a NS to be known. Furthermore, a tiny phase signal due to tidal deformation of NS (5 PN order) has to be detected. ET would be able to detect this signal, but the measurement accuracy of the redshift is not so good, $\Delta z/z\sim0.1$, and that of the luminosity distance also would be of the same order. To clarify its feasibility as a cosmological probe, further studies are needed.

LISA has marginal angular resolution so as to identify a single host galaxy. Van Den Broeck et al. \cite{bib27} considered the observation of a SMBH binary that fortunately accompanies a single host galaxy in its sky position error area. If only $w_0$ is a free parameter, the sensitivity is $\sim 0.2$. Although multiple sources improve this sensitivity, such fortunate events would be rare. To avoid this difficulty, a statistical analysis method has been developed \cite{bib88}. This method utilizes all redshifts, which obtained a priori by EM follow-up observations, of possible host galaxies in a LISA error cube and determines the best fitting cosmological model for all GW sources. Petiteau et al. \cite{bib88} performed the analysis with LISA, including the CMB prior, and obtained the accuracy of a few percents for $w_0$, though other parameters are fixed in their analysis. The applications of this statistical method to ET and DECIGO/BBO are not straightforward. The problem is that the spectroscopic catalog of galaxies is far from a complete one for a large number of sources observed by these detectors. As mentioned above, the fraction of the redshift identification of host galaxies would be $\sim10^{-4}$. Therefore, it is interesting to ask which methods are more sensitive, performing the statistical analysis with an incomplete galaxy catalog or using all GW sources without redshift information as we have done in this paper. We should address this issue in a future work.

\section{Conclusion}
\label{sec8}

\hspace*{\parindent}Proposed space-based GW detectors, DECIGO/BBO, can be a unique probe for dark energy. Using millions of NS binaries detected during the observation, we have estimated the sensitivity to the EOS parameters of the dark energy with/without identifying the redshifts of host galaxies. As a result, we found that the detectors without the redshift information have constraining power competitive to the future EM observations. This is a great advantage since the GW detector alone can probe for the cosmological expansion and enables us to compare the data of purely GW observation with those in other EM observations. With the help of the redshift information, ${\rm{FoM}}\approx 100$ corresponding to the stage III or IV of the dark energy task force is easily achieved with at most a few hundreds of sources. The standard sirens without redshifts guarantee minimally achievable FoM and strongly supports the feasibility of the space-based GW detectors as a dark energy probe. 

\section*{Acknowledgments}
A. N. and K. Y. are supported by a Grant-in-Aid through JSPS. K. Y. is also supported in part by the Grant-in-Aid for the Global COE Program from the MEXT of Japan. A. T. is 
supported in part by a Grants-in-Aid for Scientific Research from 
the JSPS No. 21740168. 

\section*{References}

\bibliographystyle{iopart-num}
\bibliography{references}

\begin{thebibliography}{10}
\bibitem{bib80}
The geometric factors arising from the nonorthogonal detector arms ($60^\circ$)
  and the sky-averaged detector response functions are incorporated into the
  detector noise curve $S_{\rm{h}}^{\rm{inst}}(f)$.
\end{thebibliography}

\end{document}